\documentclass[twocolumn,prl,superscriptaddress]{revtex4-2}
\usepackage[colorlinks=true,allcolors=blue]{hyperref}
\usepackage{dcolumn}
\usepackage{subfigure}
\usepackage{bm}
\usepackage{ifpdf,braket}
\usepackage{epstopdf}
\usepackage{amsfonts}
\usepackage{xcolor}
\usepackage{graphicx}
\usepackage{amsmath,amssymb}
\usepackage{enumerate}
\usepackage{subcaption}
\usepackage{tabularx}
\usepackage{lineno,hyperref}
\usepackage{float}
\usepackage[table]{xcolor}

\begin{document}

\title{Giant Resonances in Superconducting Films Induced by Nonsuperconducting Layers}

\author{D. Andr\'e Orna T.}
\affiliation{Laborat\'orio de Microscopia Eletr\^onica de Alta Resolu\c{c}\~ao, Centro de Caracteriza\c{c}\~ao Avan\c{c}ada para Ind\'ustria de Petr\'oleo (LaMAR/CAIPE), Universidade Federal Fluminense, Niter\'oi, 24210-346, Rio de Janeiro, Brazil}
\author{Mauro M. Doria}
\affiliation{Instituto de F\'\i sica, Universidade Federal do Rio de Janeiro, 21941-972 Rio de Janeiro, Brazil}
\author{Daniel Reyes}
\affiliation{Instituto Militar de Engenharia, Pra\c{c}a General Tibúrcio, 80, 22290-270, Praia Vermelha, Rio de Janeiro, Brazil}
\author{Isa\'\i as G. de Oliveira}
\affiliation{Departamento de F\'\i sica, Universidade Federal Rural do Rio de Janeiro, UFRRJ 23890-000, Serop\'edica-RJ, Brazil}
\author{Arkady Shanenko}
\affiliation{Centre for Quantum Metamaterials, Higher School of Economics University, Moscow 101000, Russian Federation}
\author{Alexei Vagov}
\affiliation{Centre for Quantum Metamaterials, Higher School of Economics University, Moscow 101000, Russian Federation}
\author{Y. T. Xing}
\affiliation{Laborat\'orio de Microscopia Eletr\^onica de Alta Resolu\c{c}\~ao, Centro de Caracteriza\c{c}\~ao Avan\c{c}ada para Ind\'ustria de Petr\'oleo (LaMAR/CAIPE), Universidade Federal Fluminense, Niter\'oi, 24210-346, Rio de Janeiro, Brazil}

\begin{abstract}
We find that commensurate resonances in superconducting films endowed with a $SISIS$ structure, where $S$ and $I$ stand for superconducting and insulating layers, respectively, enhance the
gap to a value three to four times the bulk gap. Such resonances rely on spatially localized quantum states that arise due to the commensurability between the total film thickness and the distance between the two insulating barriers. Our results are obtained in the context of the Bogoliubov-de Gennes equations within the Anderson approximation, applied here to Bi films, where quantum size effects are possible due to the abnormally large mean free path.
\end{abstract}

\maketitle

Parallel plates embedded in a system have long served as a paradigm for addressing fundamental questions in physics, as exemplified by the Casimir effect and the Fabry--P\'erot interferometer. The Casimir effect reveals properties of the vacuum through the imbalance between electromagnetic modes inside and outside the plates, which produces a spontaneous attraction between them~\cite{suppes96,lamoreaux05}. In the Fabry--P\'erot interferometer, the plates partially reflect light and generate resonances for selected wavelengths and separations~\cite{perot899}. Motivated by these phenomena, we consider a $SISIS$ nanostructured superconducting film, where $S$ and $I$ denote superconducting and insulating layers, respectively, and show that the interior nonsuperconducting layers give rise to a distinct resonance mechanism for enhancing the gap.

Long ago Thompson and Blatt~\cite{blatt63,thompson63} showed that the superconducting gap is enhanced in thin films by the presence of shape resonances, a concept widely used in many areas of science and associated with quasi-bound states of matter created by the presence of a potential energy barrier~\cite{heck21}.
In superconducting films the boundedness stems from the electronic confinement established by the film surfaces. 
The single-electron states become the product of two-dimensional bands and one-dimensional standing waves, associated with the degrees of freedom parallel and perpendicular to the surfaces, respectively.
Thompson and Blatt found that shape resonances cause the superconducting gap to vary with film thickness, resulting in a sawtooth-like curve whose jumps are a consequence of the entrance of parallel bands into the Debye attractive window around the Fermi surface.
They have been intensively studied theoretically~\cite{shanenko06,croitoru11,valentinis16}, but have not been extensively observed, with the exceptions of epitaxially grown Pb films in the ultrathin regime~\cite{eom06,kim10}, NiBi$_3$ films grown on a Bi substrate~\cite{doria22}, and distinct nanoparticles of Pb, Sn, and V~\cite{bose10,bose14,yang11}. 
For transverse electronic states to become bound standing waves, the electrons should not scatter while traversing the film; that is, their motion must remain coherent, meaning that the mean free path must be larger than, or at least comparable to, the total film thickness. This explains why Nb films do not show shape resonances~\cite{pinto18}.

In this Letter, we propose a novel mechanism for amplifying the superconducting gap that far surpasses the Thompson and Blatt mechanism.
In a $SISIS$ film, {\it commensurate resonant states} (CRS) arise between the two $I$ barriers whenever the ratio of the total film width, $2b$, to the distance between the centers of the insulating barriers, $2a$, is an odd integer, $b/a=l_o$.
The CRS  are special standing-wave states that are highly localized between the $I$ barriers that can be studied in the context of  the one-dimensional  Schr\"odinger equation, as previously discussed~\cite{orna25}.
They cause an enhancement of the gap that can reach three to four times the bulk gap, as shown here.

We describe the  $SISIS$ nanostructured superconducting film through the Bogoliubov-de Gennes (BdG) equations,
\begin{eqnarray} \label{bdg0}
	\left[
	\begin{array}{cc}
		\displaystyle H_0(\textbf{r})
		& \displaystyle\Delta(\textbf{r})\\
		\displaystyle\Delta^{*}(\textbf{r})
		& -\displaystyle H_0(\textbf{r})\\
	\end{array}
	\right]
	\left[
	\begin{array}{c}
		\displaystyle u_n(\textbf{r})\\
		\displaystyle v_n(\textbf{r})\\
	\end{array}
	\right] = E_n
	\left[
	\begin{array}{c}
		\displaystyle u_n(\textbf{r})\\
		\displaystyle v_n(\textbf{r})\\
	\end{array}
	\right].
	\end{eqnarray}
The $n$th level has energy $E_n$, $H_0(\mathbf r)=-\hbar^2\nabla^2/2m+V(\mathbf r)-\mu$ is the single-particle Hamiltonian, $m$ is the effective electron mass, and $\mu$ is the chemical potential of the film, determined here by the normal-state spectrum at a fixed carrier density.
The pair potential is treated at zero temperature and satisfies $\Delta(\mathbf r)=V_p\sum_n v_n^*(\mathbf r)u_n(\mathbf r)$, where $V_p>0$ is the pairing interaction.
Our analysis is done within the Anderson approximation solution of the BdG equations~\cite{anderson59,degennes99}, where the eigenfunctions $u_n(\mathbf r)$ and $v_n(\mathbf r)$ are proportional to 
the Schr\"odinger-equation wavefunction, $\Psi_n(\mathbf r)$, which satisfies 
$H_0\Psi_n=\zeta_n\Psi_n$. The eigenvalues $\zeta_n$ are single-particle energies measured with respect to $\mu$.
A set of band gaps arise in this framework, defined by
$\Delta_n\equiv \int d\textbf{r} \, \Delta(\textbf{r})|\Psi_n(\textbf{r})|^2$, which
obey self-consistent equations,
$\Delta_n = \frac{1}{2}\sum_m V_{n,m}\,\Delta_m/\sqrt{\zeta_m^2+\Delta_m^2}$,
where the pairing interaction is given by the matrix elements
$V_{n,m}=V_p\int d \textbf{r}\; |\Psi_n(\textbf{r})|^2 |\Psi_m(\textbf{r})|^2$.
\begin{figure}[H]
    \centering
    \includegraphics[width=0.4\textwidth]{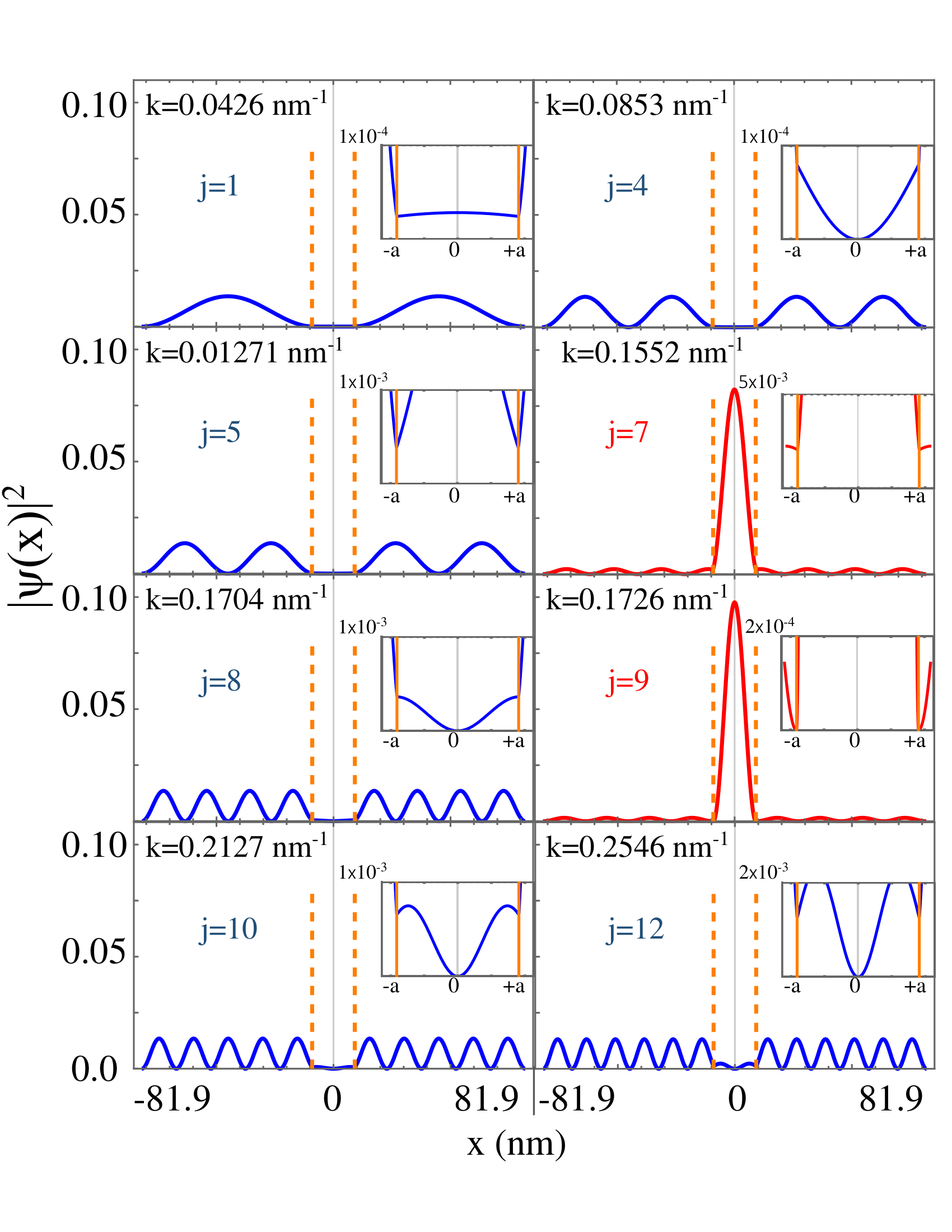}
   \caption{The probability $\vert\Psi(x)\vert^2$ is plotted versus $x$ for eight selected states from the twelve calculated states for $l_o=9$ ($2a= 18.2$, $2b= 163.8$ nm).
   The $j=1, 4, 5, 8, 10$ and $12$ states (blue online) are predominantly localized outside the barriers, while 
the CRS, $j=9$, and,  to a lesser extent the quasi-resonant state, $j=7$, are strongly confined between the barriers (red online). The insets show that only the CRS state vanishes inside the $I$ barriers.}
    \label{Fig1}
\end{figure}
For a film, we decompose real space as $\mathbf r\equiv(\mathbf r_\parallel,x)$.
Cooper pairs are confined within the film by a transverse confining potential $V(\textbf{r})=V(x)$, which describes the $SISIS$ film.
The walls are impenetrable, $V(x)=\infty$ for $\vert x \vert > b$, and the symmetrically positioned $I$  layers are  barriers of height $V(x)=V_0$ and thickness $2\epsilon$. 
They  extend over $a-\epsilon\leq |x| \leq a+\epsilon$, and so their centers are separated by $2a$. Elsewhere, the potential vanishes inside the film, $V(x)=0$.
The index $n$ becomes $n \equiv \left ( \vec k_{\parallel},k_j\right)$, associated with the continuous (parallel)  and the discrete (perpendicular) wavenumbers,  $\vec k_{\parallel}$ and $k_j$, respectively, such that the wavefunction decomposes as $\Psi_{\mathbf k_\parallel,k_{j}}(\mathbf r)=A^{-1/2}e^{i\mathbf k_\parallel\cdot\mathbf r_\parallel}\psi_{j}(x)$, where $A$ is the film area.
The functions $\psi_j(x)$ solve the  one-dimensional Schr\"odinger equation: $-(\hbar^2/2m) d^2\,\psi_j(x)/dx^2+V(x)\psi_j(x)=\varepsilon_j \psi_j(x)$.
The energy parameter becomes $\zeta_{\mathbf k_\parallel,k_j}=\varepsilon_\parallel+\varepsilon_j-\mu$, with
$\varepsilon_\parallel=\hbar^2{\mathbf k_\parallel}^2/2m$ and $\varepsilon_j=\hbar^2k_j^2/2m$. 
The bands are labeled by the transverse index $j$, and each one develops a gap, which is the projection of the spatial gap, $\Delta(x)$, onto the transverse wavefunctions,
$\Delta_j\equiv\int_{-b}^{b}dx\,\Delta(x)|\psi_j(x)|^2$.
Introducing the two-dimensional density of states per spin,
$n_{2D}=m/(2\pi\hbar^2)$, the dimensionless coupling
$\lambda \equiv k_FV_p n_{2D}/\pi$ can be obtained from the bulk gap, through
$\lambda=[\sinh^{-1}(\hbar\omega_D/\Delta_{\mathrm{bulk}})]^{-1}$~\cite{doria16,doria22,cariglia16}.
For each transverse mode $j$, $\Delta_j$ is assumed to be constant within the Debye window
$|\zeta_{\mathbf k_\parallel,j}|\le \hbar\omega_D$, where $\hbar\omega_D$ is the Debye cutoff energy; integration over the in-plane motion yields
\begin{figure}[H]
\centering
\includegraphics[width=0.5\textwidth]{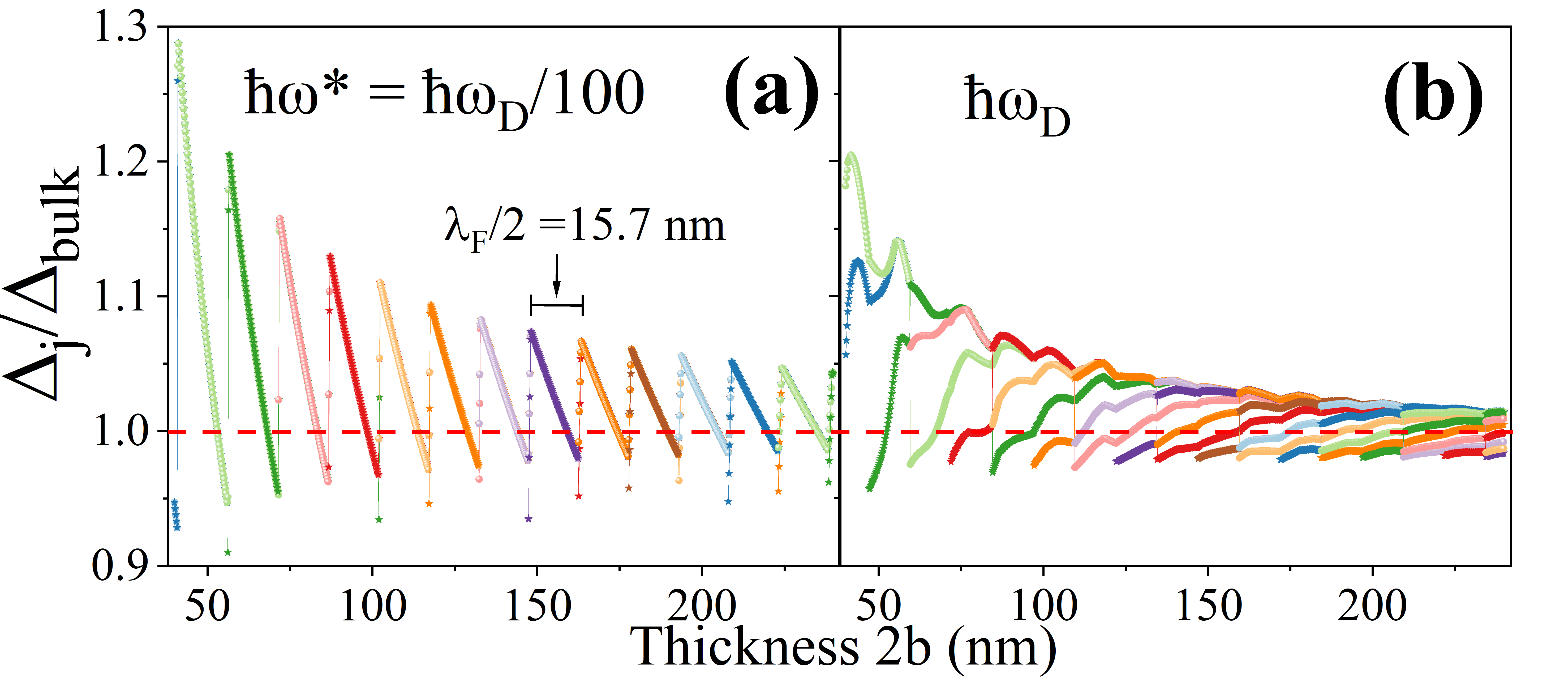}
\caption{$\Delta_j/\Delta_{bulk}$ as a function of $2b$ is shown in the absence of barriers ($V_0=0$).
Panels $(a)$ and $(b)$ refer to distinct attractive windows around $E_F$, $\hbar\omega^*=\hbar \omega_D/100$ and $\hbar\omega_{D}$, respectively.
New bands enter the film as the width $2b$ increases and are represented by different colored dots (color online).}
\label{Fig2}
\end{figure}

\begin{equation}
\hspace{-0.4cm} \Delta_j=\frac{\pi\lambda}{2k_F}\sum_i V_{ji}\,\Delta_i\,\mathcal L_i,
\, V_{ji}=\int_{-b}^{b}dx\,|\psi_j(x)|^2|\psi_i(x)|^2,
\label{eq:deltaj}
\end{equation}
where $k_F$ is the bulk Fermi wave number. The factor $\mathcal L_i$ originates from the integration over the nonnegative in-plane kinetic energy.
\begin{equation}
\mathcal L_i=\sinh^{-1}\!\left(\frac{\varepsilon_{\parallel,\max}^{(i)}+\varepsilon_i-\mu}{\Delta_i}\right)
-\sinh^{-1}\!\left(\frac{\varepsilon_{\parallel,\min}^{(i)}+\varepsilon_i-\mu}{\Delta_i}\right),
\label{eq:Li}
\end{equation}
with
$\varepsilon_{\parallel,\max}^{(i)}=\max\{0,\mu-\varepsilon_i+\hbar\omega_D\}$ and
$\varepsilon_{\parallel,\min}^{(i)}=\max\{0,\mu-\varepsilon_i-\hbar\omega_D\}$.
The gap spatial profile is therefore
\begin{equation}
\Delta(x)=\frac{\pi\lambda}{2k_F}\sum_i |\psi_i(x)|^2\,\Delta_i\,\mathcal L_i.
\label{eq:deltax}
\end{equation}
The wavefunctions $\psi_j(x)$ play a fundamental role in determining both the band gaps, $\Delta_j$, and the spatial gap, $\Delta(x)$.
This is fundamental for understanding the present mechanism, as the CRS correspond to states whose wavefunctions $\psi_j(x)$ are substantially enhanced between the $I$ barriers.
\begin{figure}[H]
\centering
\includegraphics[width=0.5\textwidth]{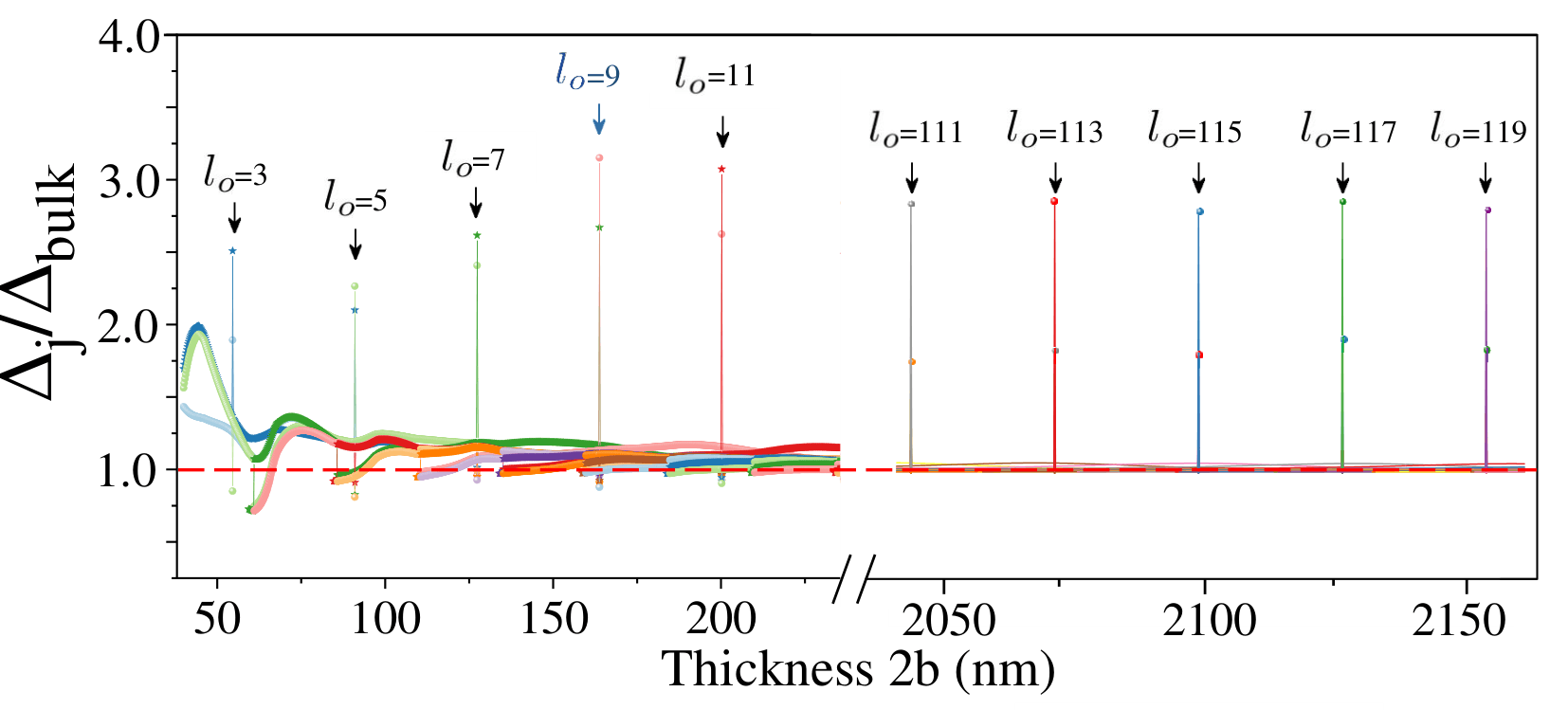}
\label{fig:sr_Y_V_156}
\caption{$\Delta_j/\Delta_{bulk}$ as a function of film thickness, $2b$, is shown here for a $SISIS$-structured film in which the $I$ layers are $I$ barriers with $V_0=625$ meV. 
Shape and commensurate resonances coexist, the latter being marked by sharp spikes associated with peaked band gaps. 
Arrows point to resonant and quasi-resonant states. 
Both the thin-film ($l_o=3$--$11$) and thick-film ($l_o=111$--$119$) limits are investigated, and in the latter, shape resonances are no longer significant as $\Delta_j /\Delta_{bulk} \sim 1$. Nevertheless, isolated peaks remain with heights $\Delta_j /\Delta_{bulk} \sim 3.0$ due to the CRS, confirming their presence in the bulk limit.}
\label{Fig3}
\end{figure}

In the limit in which the $I$ barriers are much higher than
the Fermi energy, $V_0 \gg E_F$, and the barriers are very thin, $b>a\gg\epsilon$, the standing waves $k_j$ satisfy the eigenvalue equations,
\begin{eqnarray}
\hspace{-1.0cm}&&\cos k_j b+\;\frac{2\Omega_0 }{k_j} \sin k_j(b-a)\ \cos k_j a =0
\label{eq:eigen_delta_even0} \, \mbox{(even states)}\\
\hspace{-1.0cm}&& \sin k_j b+\frac{2 \Omega_0}{k_j} \sin k_j(b-a) \sin\;k_j a =0, 
\label{eq:eigen_delta_odd0} \, \mbox{(odd states)}. \label{omega0}
\end{eqnarray}
Here, $\Omega_0 \equiv \frac{q_F}{2}\tanh(2\epsilon q_F)$, $q_F = k_F\sqrt{V_0/E_F}$.
Under the assumption that $\tanh(2\epsilon q_F)\approx 2\epsilon q_F$, the barriers act as delta-function potentials, which are reached in the limit in which $V_0\epsilon$ remains finite while $V_0 \rightarrow \infty$ and $\epsilon \rightarrow 0$.
We make this additional assumption while keeping $V_0$ finite and $\epsilon$ nonzero in our numerical calculations.
In this way, it suffices to specify the wavefunction outside the barriers.
For the even states,
\begin{equation}
\psi_j^{(e)}(x)=
\begin{cases}
A_j^{out (e)}\sin[k_j(x+b)], &-b\le x\le -a,\\
A_j^{in (e)}\cos(k_jx), &|x|\le a,\\
-A_j^{out (e)}\sin[k_j(x-b)],  &a\le x\le b,
\end{cases}
\label{eq:wave-e}
\end{equation}
where the ratio of amplitudes is $A_j^{in (e)}/A_j^{out (e)} = \sin[k_j(b-a)]/\cos(k_j a)$.
Similarly, for the odd states,
\begin{equation}
\psi_j^{(o)}(x)=
\begin{cases}
A_j^{out (o)}\sin[k_j(x+b)], & -b\le x\le -a,\\
-A_j^{in (o)}\sin(k_jx), & |x|\le a,\\
A_j^{out (o)}\sin[k_j(x-b)], & a\le x\le b,
\end{cases}
\label{eq:wave-o}
\end{equation}
where $A_j^{in (o)}/A_j^{out (o)}=\sin[k_j(b-a)]/\sin(k_j a)$.

To understand the CRS, consider the well-known infinite-well eigenvalue problem ($V(x) = 0$ for $\vert x \vert \le a$ and $V(x)=\infty$ for $\vert x \vert > a$), whose 
states satisfy $\cos{k_ja} = 0$ (even) and $\sin{k_ja} = 0$ (odd).
Interestingly, such states exist in the $SISIS$ film, provided that the commensurate condition $b/a=l_o$ holds~\cite{orna25}, according to Eqs.~\eqref{eq:eigen_delta_even0} and \eqref{eq:eigen_delta_odd0}. 
Not all states under the commensurate condition are CRS, but the reverse is true, as can be checked in Eqs.~\eqref{eq:wave-e} and \eqref{eq:wave-o}.
Remarkably, the CRS are concentrated between the barriers~\cite{orna25}, as quantified by the squared probability-amplitude ratio 
\begin{eqnarray}
\rho \equiv \left(\frac{A_j^{in (e)}}{A_j^{out (e)}}\right)^2 = \left(\frac{A_j^{in (o)}}{A_j^{out (o)}}\right)^2= (l_o-1)^2,
\label{rho}
\end{eqnarray}
which is equal for both parities only for the CRS, and  very large for $l_o \gg 1$.
To illustrate the present ideas about the CRS, we choose as the conducting medium the only superconducting phase of Bi without applied pressure~\cite{valladares18,khasanov19}.  
Bi is a semimetal and has both electrons and holes~\cite{brice18}, but in the present simplified model only  electrons are considered.
Its properties have been studied in both bulk samples and films ~\cite{barret60,wolff64,partin89,meilu96, chou22}.
The Bi mean free path is abnormally large, allowing electrons to remain coherently confined in transverse standing states in the film: $\xi_s\sim 0.38$ $\mu$m and $1.0$ $\mu$m for films~\cite{hulst95} and bulk samples~\cite{sondheimer52,pippard52}, respectively.
The  electronic density~\cite{hartman69,prakash17,koley17} is very low, $3.0 \times 10^{-4}$ nm$^{-3}$, which yields a Fermi wave number $k_F= 0.2$ nm$^{-1}$ and a Fermi wavelength  $\lambda_F = 31.4$ nm~\cite{hulst95, brice18}.
The Fermi energy is $E_F = 25$ meV~\cite{wolff64,garcia72,mikhail81,alstrom81,prakash17}, and  from $E_F=\hbar^2 k_F^2/2m$, it follows that the electronic effective mass is $m = 0.065 \; m_e$, where $m_e$ is the electron's mass. For simplicity, we disregard the tensorial nature of the mass~\cite{sedov17} and the value obtained is basically the mass along the film surface~\cite{hulst95}. 
The Debye energy of Bi $\hbar\omega_D=12$ meV has long been known~\cite{barret60}.
About a decade ago, O. Prakash  \textit{et al.} ~\cite{prakash17} found superconductivity in the millikelvin regime for this Bi phase, and 
measured its critical field~\cite{prakash17}, from which the zero-temperature gap is obtained as $\Delta_{bulk}(0)=3.9$ meV, 
by first determining the density of states at the Fermi level from $N_F=m k_F/2\pi^2 \hbar^2$.
This gives the ratio $\Delta_{bulk}(0)/\hbar \omega_D = 0.327$, which yields the dimensionless coupling  $\lambda=0.544$.
Note that  $\hbar \omega_D/E_F= 0.48$, indicating that a large fraction of electrons participate in the superconducting state, causing a profound deformation of the Fermi surface in the superconducting state.

The presence of CRS strongly enhances the band gaps,  $\Delta_j$, and the spatial gap, $\Delta(x)$, according to Eqs.~\eqref{eq:deltaj} and \eqref{eq:deltax}.
In what follows we study this enhancement for  particular $I$ barriers with height $V_0=625$ meV, thickness $2\epsilon=1.0$ nm, and separation $2a=18.2$ nm.
Note that in this case,  Eqs.~\eqref{eq:eigen_delta_even0} and \eqref{eq:eigen_delta_odd0} apply, since $V_0/E_F=25$ and $a/\epsilon=18.2$.

Fig.~\ref{Fig1} shows some transverse states derived from the one-dimensional  Schr\"odinger equation~\cite{orna25, orna26} for the special case $l_o=9$ ($2b= 163.8$ nm).
Among the possible states, one is quasi-resonant, $k_7a\sim \pi/2$, and another is resonant, $k_9a=\pi/2$, $a=9.1$ nm.
Note that the resonant state reaches the extremely high ratio of $\rho=64$, according to Eq.~\eqref{rho}.
The  matrix element $V_{i, j}$, defined in Eq.~\eqref{eq:deltaj}, reaches its maximum value  for the resonant state, $2bV_{9,9} = 10.67$, nearly ten times larger than the other matrix elements.
Theoretical formulas are in good agreement with the obtained numerical values: besides Eq.~\eqref{rho}, one obtains  $2b V_{i,j}=3(l_o-1)^2/2l_o$ for the CRS~\cite{orna26}.

Fig.~\ref{Fig2} shows the case in which CRS are absent but shape resonances are present,  obtained for no $I$ barriers ($V_0=0$).
The elevated ratio $\hbar \omega_D/E_F$ of Bi makes several bands enter simultaneously into the pairing window $\pm \hbar \omega_D$ around the chemical potential $\mu$.
This smooths out the well-known sawtooth form of the $\Delta_j/\Delta_{bulk}$ as a function of $2b$~\cite{blatt63, thompson63}.
For this reason, we have considered a fictitious cutoff, a hundred times smaller, $\hbar\omega^*= \hbar\omega_D/100$, where bands enter the attractive window one by one, rendering the onset of shape resonances easily observable.
This is displayed in~Fig.\ref{Fig2}$(a)$, where the sequence of tooth-shaped curves is seen, linked to the entrance of individual bands $j$. 
As the film width, $2b$, increases, new gaps $\Delta_j/\Delta_{bulk}$ arise due to the corresponding bands that enter the attractive window.
A periodicity of $2b \rightarrow 2b+ \lambda_F/2$ is observed, resulting in the traditional $\Delta_j/\Delta_{bulk}$ vs $2b$ sawtooth curve.
Fig.~\ref{Fig2}$(b)$ shows the band gaps obtained using the true Debye cutoff, $\hbar\omega_{D}$. 
However, the curves do not attain the sawtooth form because of the broad attractive Debye window that deforms the Fermi surface, causing overlapping contributions for the gaps associated with the individual bands, a fact that has been previously observed in Ref.~\cite{cariglia16}.
For both Panels, the chemical potential $\mu$ changes as a function of $2b$ to keep the electronic density constant, as previously discussed.

Fig.~\ref{Fig3} shows the onset of CRS amid the shape resonances in the case of a $SISIS$ structure.
The smooth sawtooth structure seen in Fig.~\ref{Fig2}$(a)$, a remnant of shape resonances, coexists with pronounced spikes.
A substantial enhancement of $\Delta_j/\Delta_{bulk}$ is observed every time the commensurability condition $b/a = l_o$ is satisfied, and this happens for $l_o =7,\,9,\,11$.
The CRS highlighted in Fig.~\ref{Fig1} play a central role in this enhancement,  according to Eq.~\eqref{eq:eigen_delta_even0}.
It is well known that shape resonances disappear as the film thickness becomes sufficiently large, turning the film into the bulk: $\Delta_j \rightarrow \Delta_{bulk}$ for $2b \rightarrow \infty$.
However, the effects of CRS persist in this limit, as they  stem from localized states that exist between the barriers and practically vanish outside them.
This enhancement of the gap persists even in the thick film limit, as shown here.

\begin{figure}[H] 
    \centering
    \includegraphics[width=0.4\textwidth]{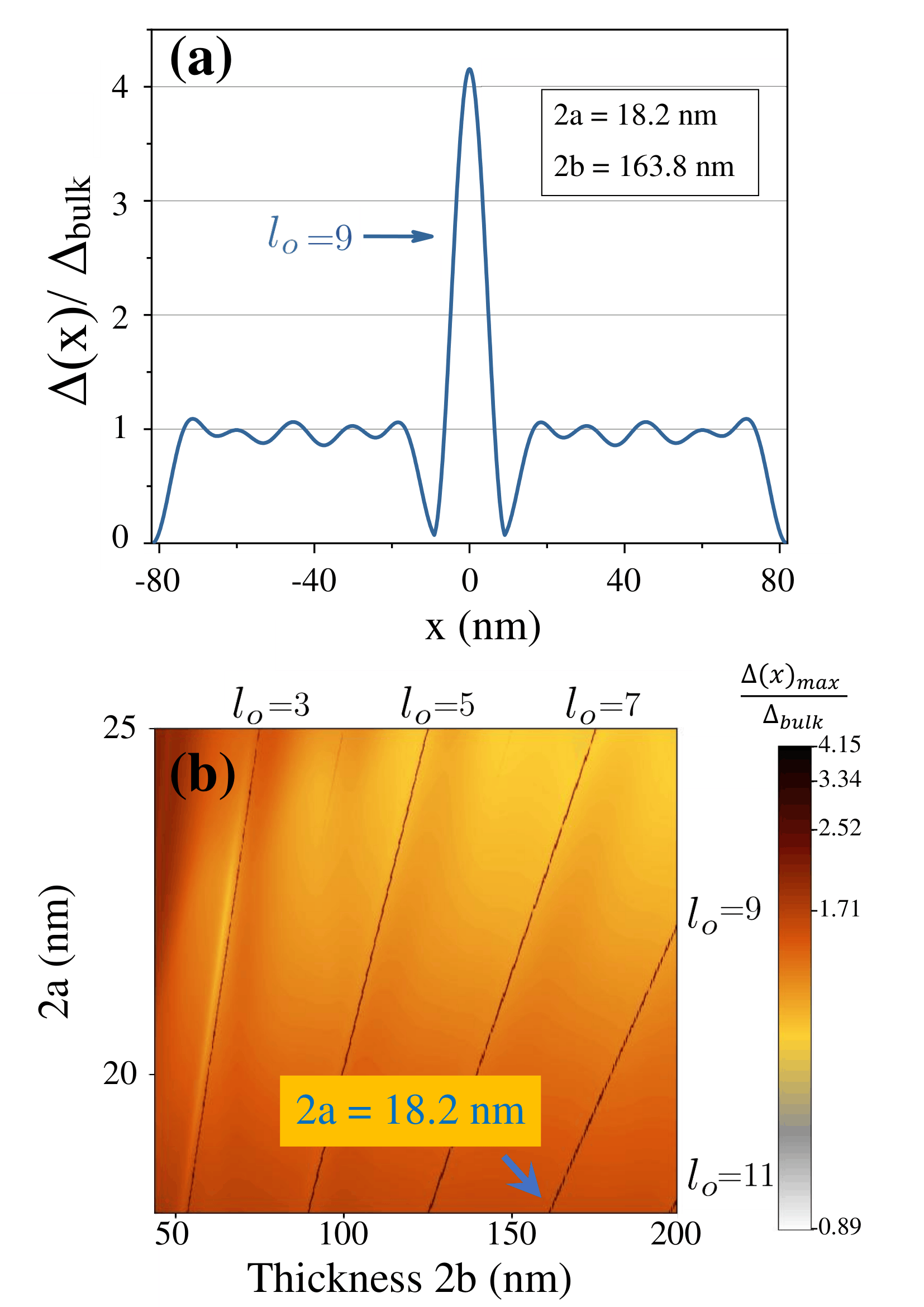}
    \caption{Panel $(a)$ shows the transverse profile of the superconducting gap,  $\Delta(x)/\Delta_{bulk}$ vs $x$, for a CRS, revealing the presence of a maximum gap.
  A search for this maximum gap is carried out in Panel $(b)$, for each pair $(a,b)$. Its magnitude is represented by a color scale (color online).
    Straight lines are clearly visible whenever the condition  $b/a=l_o$ is satisfied, indicating the presence of CRS in the film.
    The CRS parameter set of Panel $(a)$ is highlighted in Panel $(b)$ and falls on the $l_o=9$ line.}
    \label{Fig4}
\end{figure}

A thorough search for CRS is carried out in Fig.~\ref{Fig4} by sweeping the set $(a,b)$. 
The distance between the barriers and the film thickness are varied simultaneously, and for each pair, the normalized gap, $\Delta(x)/\Delta_{bulk}$, is obtained, as shown in Fig.~\ref{Fig4}$(a)$. 
The gap is found to reach a maximum between the $I$ barriers nearly four times the bulk value.
This enhancement is a consequence of the CRS, shown in Fig.~\ref{Fig1}, which affect $\Delta(x)$, according to Eq.~\eqref{eq:deltax}.
In summary, outside the barriers the gap is that of the bulk, inside them the gap vanishes, and between them it is enhanced. 
Fig.~\ref{Fig4}$(b)$ shows this maximum gap for each pair $(a,b)$ through a color map, where
a series of bright yellow (color online) dotted lines are observable, clearly distinguishable over the smooth purple (color online) background in this figure. 
They are  straight-line trajectories corresponding to the odd–integer commensurability condition $b = l_{o}a$,  visible for $l_{o}=3,\,5,\,7,\,9$, and $11$. 

In conclusion, we have demonstrated here a powerful mechanism for enhancing the superconducting gap in films by endowing them with a $SISIS$ structure.
The gap enhancement far surpasses that of the shape-resonance mechanism of Thompson and Blatt~\cite{blatt63, thompson63}.
\begin{acknowledgments}
We are grateful to Robert Marino Espinoza for helpful and illuminating discussions. 
DAOT acknowledges financial support for a doctoral scholarship from FAPERJ with grant No. E-26/210.742/2025 and PETROBRAS with grant No. 0050.0129649.24.9.
DR, MMD, and IGdO acknowledge support from the INCT project Advanced Quantum Materials, involving the Brazilian agencies CNPq (Proc. 408766/2024-7), FAPESP, and CAPES. MMD, AS, and AV acknowledge support from the project ``International academic cooperation'' of the Higher School of Economics University (HSE) of the Russian Federation. DR would like to thank the Aux\'ilio B\'asico \`a Pesquisa (APQ1), Grant No. SEI-260003/006384/2024, Ref.: 210.969/2024 E13/2023 from Funda\c{c}\~ao de Amparo \`a Pesquisa do Estado do Rio de Janeiro (FAPERJ), and the Peruvian Agency CONCYTEC for financial support through Grant No. PE501096507-2025-PROCIENCIA.
\end{acknowledgments}

\bibliography{reference}
\end{document}